\documentclass[journal=ancac3,manuscript=letter]{achemso}
\usepackage[T1]{fontenc}
\usepackage[latin1]{inputenc}
\usepackage{lmodern}\usepackage{graphicx}
\usepackage{dcolumn}
\usepackage{bm}
\usepackage{amsmath}
\usepackage{amssymb}
\usepackage{pst-all}
\usepackage{psfrag}
\usepackage{epsfig}

\def\ii{{\rm i}}  \def\ee{{\rm e}}  \def\vb{{\bf v}}
\def\Rb{{\bf R}}      \def\ub{{\bf u}}
\def\jb{{\bf j}}  \def\kb{{\bf k}}    \def\rb{{\bf r}}
    
\def\th{\vec{\bf{\theta}}} 
\def\fE{\vec{\mathcal{E}}}  \def\Mb{{\bf M}}

\title{Multiple Excitation of Confined Graphene Plasmons by Single Free Electrons}
\author{F.~Javier~Garc\'{\i}a~de~Abajo}
\affiliation[ICFO]{ICFO - Institut de Ciencies Fotoniques, Mediterranean Technology Park, 08860 Castelldefels (Barcelona), Spain}
\altaffiliation{ICREA - Instituci\'o Catalana de Recerca i Estudis Avan\c{c}ats, Barcelona, Spain}
\email{javier.garciadeabajo@icfo.es}

\begin{document}

\begin{abstract}
We show that free electrons can efficiently excite plasmons in doped graphene with probabilities of order one per electron. More precisely, we predict multiple excitations of a single confined plasmon mode in graphene nanostructures. These unprecedentedly large electron-plasmon couplings are explained using a simple scaling law and further investigated through a general quantum description of the electron-plasmon interaction. From a fundamental viewpoint, multiple plasmon excitations by a single electron provides a unique tool for exploring the bosonic quantum nature of these collective modes. Our study does not only open a viable path towards multiple excitation of a single plasmon mode by single electrons, but it also reveals graphene nanostructures as ideal systems for producing, detecting, and manipulating plasmons using electron probes.
\end{abstract}
\maketitle


The existence of surface plasmons was first demonstrated by observing energy losses produced in their interaction with free electrons \cite{R1957,PS1959}. Following those pioneering studies, electron beams have revealed many of the properties of plasmons through energy-loss and cathodoluminescence spectroscopies, which benefit from the impressive combination of high spatial and spectral resolutions that is currently available in electron microscopes and that allows mapping plasmon modes in metallic nanoparticles and other nanostructures of practical interest \cite{paper085,RCV11,paper149}. However, plasmon creation rates are generally low, thus rendering multiple excitations of a single plasmon mode by a single electron extremely unlikely. 

From a fundamental viewpoint, the question arises, what it the maximum excitation probability of a plasmon by a passing electron? This depends on a number of parameters, such as the interplay between momentum and energy conservation during the exchange with the electron, the spatial extension of the  electromagnetic fields associated with the electron and the plasmon, and the interaction time, which are in turn controlled by the spatial extension of the excitation and the speed of the electron. One expects that highly confined optical modes, encompassing a large density of electromagnetic energy, combined with low-energy electrons, which experience long interaction times, provide an optimum answer to this question. This intuition is corroborated here by examining the interaction between free electrons and graphene plasmons. The peculiar electronic structure of this material leads to the emergence of strongly confined plasmons (size $<1/100$ of the light wavelength) when the carbon layer is doped with charge carriers \cite{CGP09,JBS09,VE11,paper176,NGG11_2}. Graphene plasmons have been recently observed and their electrical modulation unambiguously demonstrated through near-field spatial mapping \cite{paper196,FRA12,YLZ12} and far-field spectroscopy \cite{JGH11,paper212,BJS13}. These low-energy plasmons, which appear at mid-infrared and lower frequencies, should not be confused with the higher-energy $\pi$ and $\sigma$ plasmons that show up in most carbon allotropes, and that have been extensively studied through electron energy-loss spectroscopy (EELS) in fullerenes \cite{KC92,LMG13}, nanotubes \cite{STK02}, and graphene \cite{EBN08,ZLN12,DDS12,CLC13}. These high-energy plasmons are not electrically tunable. We thus concentrate on electrically driven low-energy plasmons in graphene. Despite their potential for quantum optics and light modulation \cite{HNG12,paper184,paper204}, the small size of the graphene structures relative to the light wavelength (see below) poses the challenge of controlling their excitation and detection with suitably fine spatial precision. Using currently available subnanometer-sized beam spots, free electrons appear to be a viable solution to create and detect graphene plasmons with large yield and high spatial resolution. As a first step in this direction, angle-resolved EELS performed with low-energy electrons has been used to map the dispersion relation of low-energy graphene plasmons \cite{LWE08,TPL11}, as well as their hybridization with the phonons of a SiC substrate \cite{LW10}, although this technique has limited spatial resolution.

The probabilities of multiple plasmon losses, as observed in EELS \cite{SFS1987} and photoemission \cite{AHK96} experiments, are well known to follow Poisson distributions \cite{SL1971,MSG10,LMG13}. Previous studies have concentrated on plasmon bands, where the electrons simultaneously interact with a large number of plasmon modes. We are instead interested in the interaction with a spectrally isolated single mode.

In this article, we show that a single electron can generate graphene plasmons with large yield of order one. We discuss the excitation of both propagating plasmons in extended carbon sheets and localized plasmons in nanostructured graphene. The excitation probability is shown to reach a maximum value when the interaction time is of the order of the plasmon period. Our results suggest practical schemes for the excitation of multiple localized graphene plasmons using electron beams, thus opening new perspectives for the observation of nonlinear phenomena at the level of a few plasmons excited within a single mode of an individual graphene structure by a single electron.

\begin{figure}
\begin{center}
\includegraphics[width=85mm,angle=0,clip]{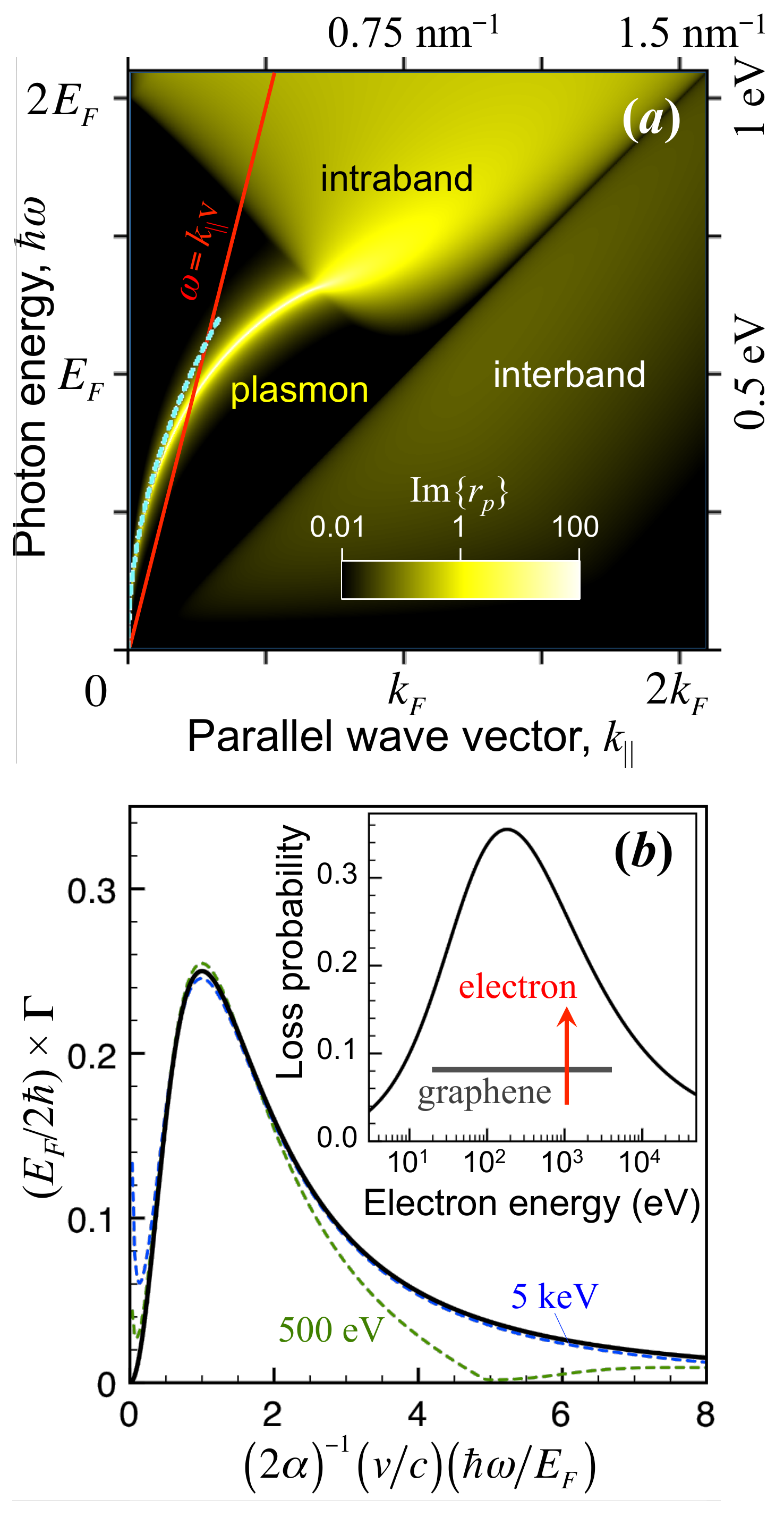}
\caption{{\bf Excitation of graphene plasmons by electron beams.} {\bf (a)} Frequency and parallel-wave-vector dependence of the Fresnel reflection coefficient $r_p$ for $p$ polarized light, calculated in the random-phase approximation (RPA) and showing plasmons and interband/intraband electron-hole-pair transitions in doped extended graphene. The Fermi energy and intrinsic damping time are $E_F=0.5\,$eV and $\tau=0.5\,$ps, respectively. An electron moving with velocity $v$ couples preferentially to excitations in the $k_\parallel v\sim\omega$ region. {\bf (b)} Electron energy-loss spectra for an electron incident normally to a homogeneous graphene sheet under the same conditions as in (a) for two different electron kinetic energies (broken curves), compared with the universal analytical expression for the plasmon contribution derived in the Drude model (solid curve, \ref{Gperpana}). The inset shows the integral of the latter over the $\hbar\omega<E_F$ region as a function of electron energy.} \label{Fig1}
\end{center}
\end{figure}

\section{RESULTS AND DISCUSSION}

{\bf Plasmon excitation in extended graphene.} Before discussing confined plasmons, we explore analytical limits for electrons interacting with a homogeneous graphene layer. The dispersion relation of free-standing graphene plasmons can be directly obtained from the pole of the Fresnel coefficient for $p$ polarization \cite{JBS09}, which in the electrostatic limit reduces to
\begin{equation}
r_p=\frac{1}{1-\ii\omega/2\pi k_\parallel\sigma}.
\label{rp}
\end{equation}
Here, $\sigma(k_\parallel,\omega)$ is the conductivity, and $k_\parallel$ and $\omega$ are the light parallel wave-vector and frequency. Because of the translational invariance of the carbon layer, the full $k_\parallel$ dependence of the conductivity $\sigma$ can be directly incorporated in \ref{rp} to account for nonlocal effects, which include the excitation of electron-hole pairs. The dispersion diagram of \ref{Fig1}a, which shows ${\rm Im}\{r_p\}$ calculated in the random-phase approximation \cite{WSS06,HD07} (RPA) under realistic doping conditions (Fermi energy $E_F=0.5\,$eV, corresponding to a carrier density $n=(E_F/\hbar v_F)^2/\pi=1.84\times10^{13}\,$cm$^{-2}$, where $v_F\approx10^6\,$m/s is the Fermi velocity), reveals a plasmon band as a sharp feature outside the regions occupied by intra- and interband electron-hole-pair transitions. We justify the use of the electrostatic approximation to describe the response of graphene because the plasmon wavelength is much smaller than the light wavelength ({\it e.g.}, the Fermi wavelength is $2\pi/k_F=\sqrt{4\pi/n}=8.3\,$nm, which is 300 times smaller than the light wavelength at an energy $\hbar\omega=E_F=0.5\,$eV).

Under these conditions, an electron crossing the carbon sheet with constant normal velocity $v$ has a probability \cite{paper149}
\begin{equation}
\Gamma_\perp(\omega)=\frac{4e^2}{\pi\hbar v^2}\int_0^\infty\frac{k_\parallel^2\,dk_\parallel}{\big(k_\parallel^2+\omega^2/v^2\big)^2}{\rm Im}\left\{r_p\right\} \label{Gperp}
\end{equation}
of lossing energy $\hbar\omega$ (see Methods for more details). Likewise, an electron moving along a path length $L$ parallel to the graphene experiences a loss probability (see Methods)
\begin{equation}
\Gamma_\parallel(\omega)=\frac{2e^2L}{\pi\hbar v^2}\int_{\frac{\omega}{v}}^\infty\frac{dk_\parallel}{\sqrt{k_\parallel^2-\omega^2/v^2}}\,\ee^{-2k_\parallel z_0}\,{\rm Im}\left\{r_p\right\}, \label{Gpara}
\end{equation}
where $z_0$ is the distance between graphene and the electron. These probabilities are determined by ${\rm Im}\left\{r_p\right\}$, which is represented in \ref{Fig1}a. Clearly, losses around the $k_\parallel\sim\omega/v$ region are favored.

It is instructive to evaluate \ref{Gperp,Gpara} using the Drude model for the graphene conductivity \cite{JBS09}
\begin{equation}
\sigma(\omega)=\frac{e^2E_F}{\pi\hbar^2}\frac{\ii}{\omega+\ii\tau^{-1}},
\label{Drude}
\end{equation}
where $\tau$ is a phenomenological decay time. The latter determines the plasmon quality factor $Q=\omega\tau\sim10-60$, as measured in recent experiments \cite{FRA12,paper196,paper212,BJS13,YLZ12}. This model works well for photon energies below the Fermi level ($\hbar\omega<E_F$), but neglects interband transitions that take place at higher energies. In the $\omega\tau\gg1$ limit, we can approximate 
\begin{equation}
{\rm Im}\left\{r_p\right\}\approx\pi k_\parallel^{\rm SP}\delta(k_\parallel-k_\parallel^{\rm SP}),
\nonumber
\end{equation}
where
\begin{equation}
k_\parallel^{\rm SP}=(\hbar^2\omega^2/2e^2E_F)
\label{kSP}
\end{equation}
is the plasmon wave vector (broken curve in \ref{Fig1}a). Using this approximation in \ref{Gperp,Gpara}, the plasmon contribution to the loss probability reduces to
\begin{equation}
\Gamma_\perp(\omega)\approx\frac{2\hbar}{E_F} \frac{s^2}{\big(1+s^2\big)^2}, \label{Gperpana}
\end{equation}
\begin{equation}
\Gamma_\parallel(\omega)\approx\frac{\hbar\omega L}{vE_F}\,\frac{\theta(s-1)}{\sqrt{s^2-1}}\,\ee^{-k_\parallel^{\rm SP}z_0}, \nonumber
\end{equation}
where
\begin{equation}
s=\frac{1}{2\alpha}\frac{v}{c}\frac{\hbar\omega}{E_F} \nonumber
\end{equation}
and $\alpha\approx1/137$ is the fine-structure constant. For a perpendicular trajectory, \ref{Gperpana} has a maximum at $s=1$ (\ref{Fig1}b). Its integral over the $\hbar\omega<E_F$ region, in which the plasmon is well defined, shows a maximum probability $>35\%$ for an electron energy $\sim215\,$eV when we take $E_F=0.5\,$eV. This is an unusually high plasmon yield for electrons traversing a thin film. However, this probability is spread over a continuum of 2D plasmons. In what follows, we concentrate on localized plasmons in finite graphene islands, which feature instead a discrete spectrum, thus placing the entire electron-plasmon strength on a few modes, and consequently, increasing the probability for a single electron to excite more than one plasmon in a single mode.

\begin{figure}
\begin{center}
\includegraphics[width=140mm,angle=0,clip]{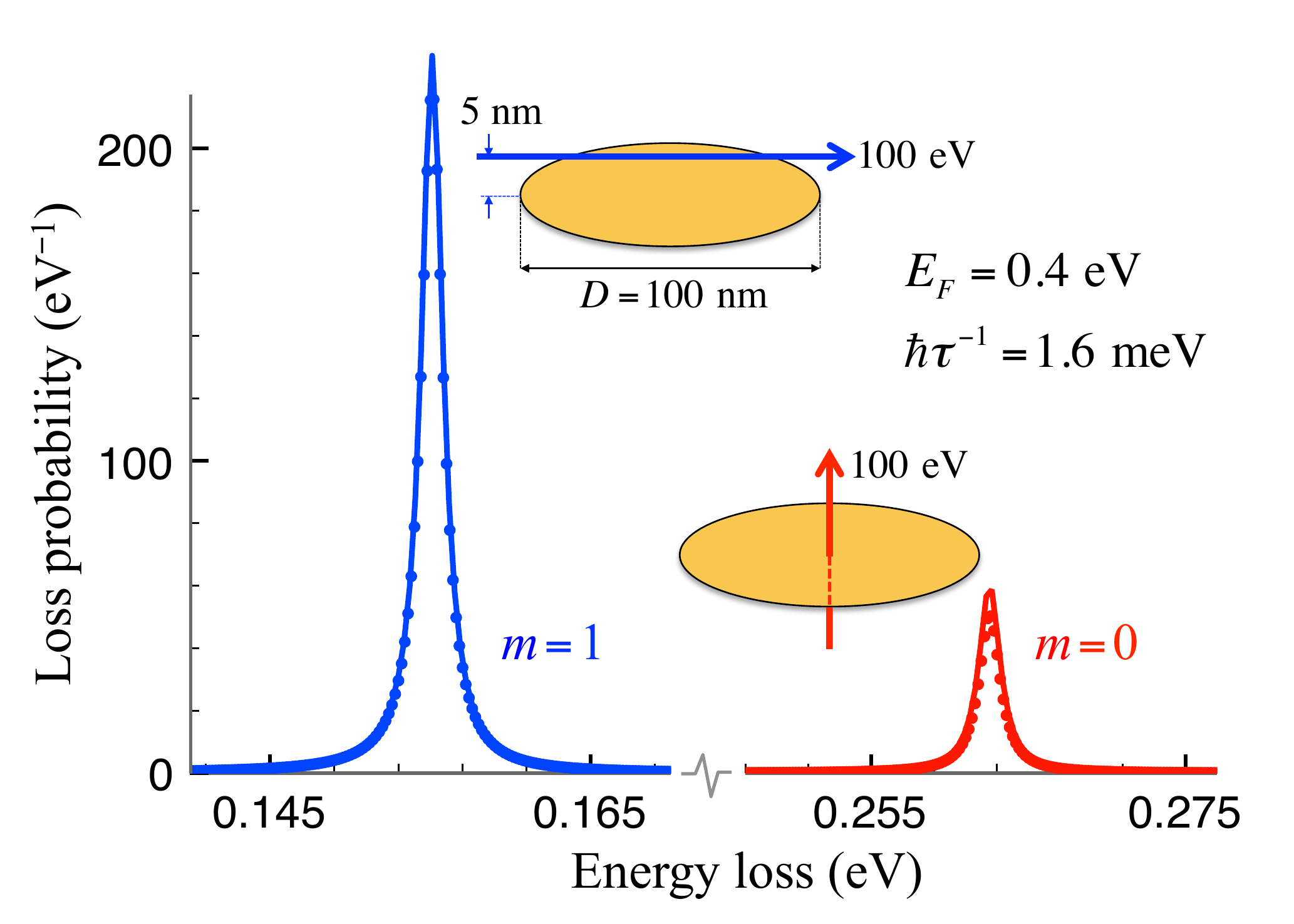}
\caption{{\bf Plasmons excitation in a graphene nanodisk.} Electron energy-loss probability for two different electron trajectories exciting plasmon modes of different azimuthal symmetry $m$, as shown by labels.} \label{Fig2}
\end{center}
\end{figure}

{\bf Plasmon excitation in graphene nanostructures.} Doped graphene nanoislands can support plasmons, as recently observed through optical absorption measurements \cite{paper212,BJS13}. For small islands, the concentration of electromagnetic energy that characterizes these plasmons is extremely high, therefore producing strong coupling with nearby quantum emitters \cite{paper176}. Likewise, the interaction of localized plasmons with a passing electron is expected to be particularly intense. This intuition is put to the test in \ref{Fig2}, where we consider a 100\,eV electron interacting with a 100\,nm graphene disk doped to a Fermi energy $E_F=0.4$\,eV. We calculate the electron energy-loss probability using classical electrodynamics \cite{paper149}, assuming linear response, and describing the graphene through the Drude conductivity of \ref{Drude}, which we spread over a thin layer of $\sim0.3\,$nm thickness, as prescribed elsewhere \cite{paper176}. We investigate both a parallel trajectory, with the electron passing 5\,nm away from the carbon sheet, and a perpendicular trajectory, with the electron crossing the disk center. \ref{Fig2} only shows the lowest-energy mode that is excited for each of these geometries, corresponding to $m=1$ and $m=0$ azimuthal symmetries, respectively. An overview of a broader spectral range (see Methods, \ref{FigS2}) reveals that these modes are actually well isolated from higher-order spectral features within each respective symmetry. Because we use the Drude model, the width of the plasmon peaks in \ref{Fig2} coincides with $\hbar\tau^{-1}$ (see below), which is set to 1.6\,meV, or equivalently, we consider a mobility of $10,000\,$cm$^2/($V\,s$)$, which is a moderate value below those measured in suspended \cite{BSJ08} and BN-supported \cite{DYM10} high-quality graphene.

The area of the plasmon peaks is a $\tau$-independent, dimensionless quantity that corresponds to the number of plasmons excited per incident electron. For the $m=1$ mode (parallel trajectory), we find $\sim0.4$ plasmons per electron. This leads us to the following two important conclusions: (i) the probability of multiple plasmon generation is expected to be significant, and its study requires a quantum treatment of the plasmons to cope with the bosonic statistics of these modes, as described below; and (ii) linear response theory, which we assume within the classical electromagnetic calculations shown in \ref{Fig2}, is no longer valid because nonlinear and quantum corrections become substantial. These two conclusions are further explored in what follows, but first we discuss a universal scaling law for the energy-loss probability that is also relevant to the description of multiple plasmon processes.

{\bf Electrostatic scaling law and maximum plasmon excitation rate.} In electrodynamics, the light wavelength introduces a length scale that renders the solution of specific geometries size dependent. In contrast, electrostatics admits scale-invariant solutions, which have long been recognized to provide convenient mode decompositions, particularly when studying electron energy losses \cite{OI1989,paper010,BK12}. Modeling graphene as an infinitely thin layer, its electrostatic solutions take a particularly simple form \cite{paper212}. We provide a comprehensive derivation of the resulting scaling laws in the Methods section, the main results of which are summarized next. We focus on a spectrally isolated plasmon of frequency $\omega_p$ sustained by a graphene nanoisland of characteristic size $D$ ({\it e.g.}, the diameter of a disk) and homogeneous Fermi energy $E_F$.

Assuming the Drude model for the graphene conductivity (\ref{Drude}), the plasmon frequency is found to be
\begin{equation}
\omega_p=\gamma_p\frac{e}{\hbar}\sqrt{\frac{E_F}{D}},
\label{wplast}
\end{equation}
where $\gamma_p$ is a dimensionless, scale-invariant parameter that only depends on the nanoisland geometry and plasmon symmetry under consideration (see \ref{gamp}). In particular, we have $\gamma_p=3.6$ for the lowest-order axially symmetric plasmon of a graphene disk ($m=0$ azimuthal symmetry), which is the lowest-frequency mode excited by an electron moving along the disk axis. Also, we find $\gamma_p=2.0$ for the lowest-order $m=1$ mode, which can be excited in asymmetric configurations. It is important to stress that \ref{wplast} reveals a linear dependence of $\omega_p$ on $\sqrt{E_F/D}$.

Furthermore, the average number of plasmons excited by the electron ({\it i.e.}, the plasmon yield) reduces to (see Methods, \ref{Gj,Pj})
\begin{equation}
P_p^{\rm cla}\approx\frac{1}{\mu}\;\;F_p\!\left(\frac{\mu}{\nu}\right)
\label{PF}
\end{equation}
within this classical theory. Here, we have defined the two dimensionless parameters
\begin{subequations}
\begin{equation}
\mu=\frac{\sqrt{E_FD}}{e}=\sqrt{\frac{E_FD}{\alpha\hbar c}},
\end{equation}
\begin{equation}
\nu=\frac{\hbar v}{e^2}=\frac{v}{\alpha c},
\end{equation}
\label{munu}
\end{subequations}
as well as the dimensionless loss function
\begin{equation}
F_p\left(x\right)=x^2\left|\;\int d\ell\;\exp\!\left(\ii\gamma_px\ell\right)\;f_p(\ell)\right|^2.
\label{Fpxi}
\end{equation}
The integral is over the electron path length $\ell$ in units of $D$, whereas $f_p$ is the dimensionless scaled plasmon electric potential defined in \ref{fj}, which is calculated once and for all following the procedure explained in the Methods section (see \ref{PpQ} and beyond).

\begin{figure}
\begin{center}
\includegraphics[width=100mm,angle=0,clip]{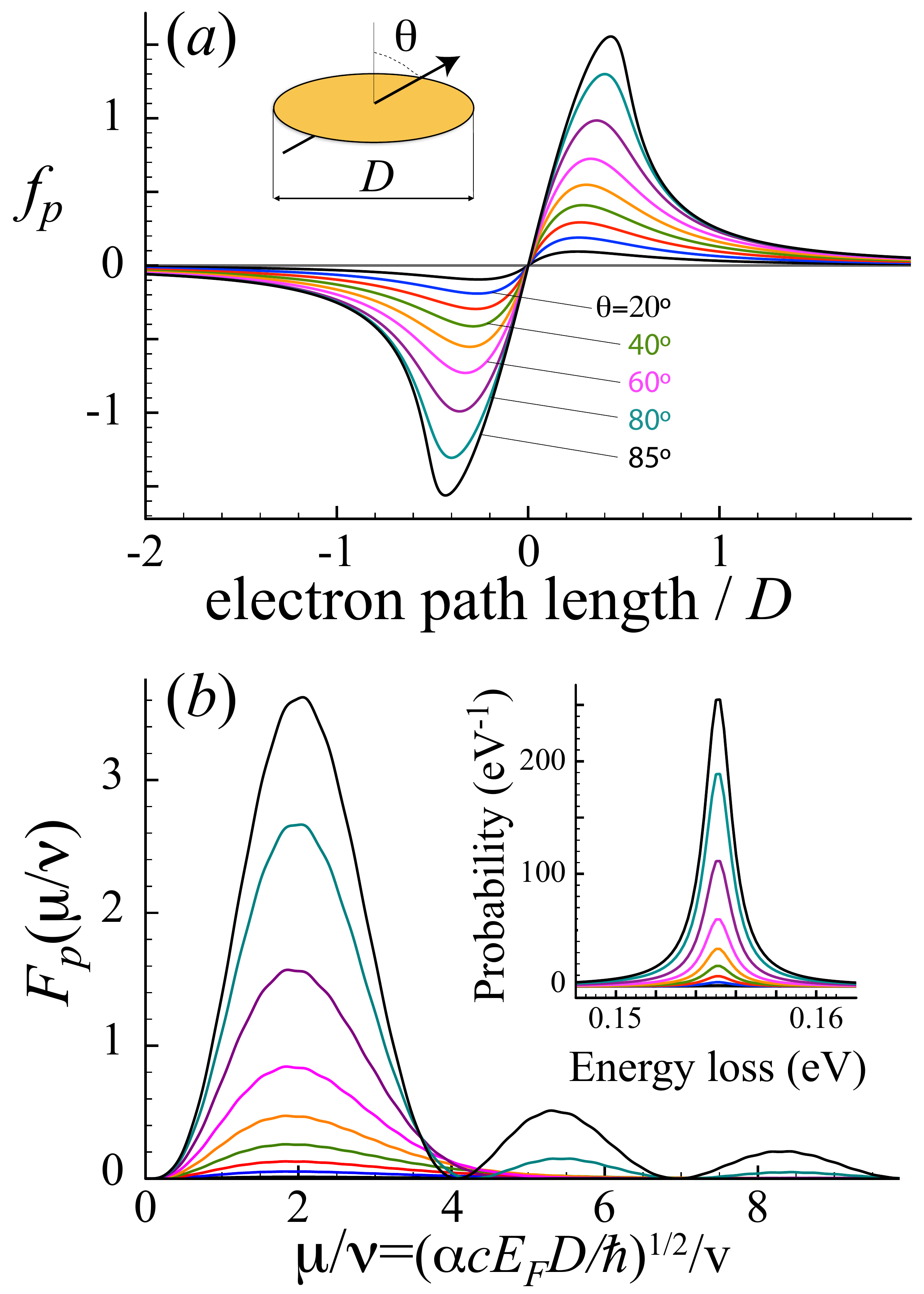}
\caption{{\bf Scaled potential and loss function.} {\bf (a)} Scaled plasmon electric potential $f_p$ (see Methods, \ref{fj})  sampled by different electron trajectories crossing the center of a graphene disk with different angles relative to the graphene normal, as shown by labels. {\bf (b)} Scaled loss function $F_p$ for the lowest-oder $m=1$ plasmon excited under the trajectories considered in (a). The inset shows the energy-loss probability for a plasmon width of 1.6\,meV.} \label{Fig3}
\end{center}
\end{figure}

We show characteristic examples of $f_p$ and $F_p$ in \ref{Fig3} for electrons crossing the center of a graphene disk following different oblique trajectories. As $f_p$ is proportional to the electrostatic potential associated with the plasmon, it is a real function of position. For the $m=1$ mode considered in \ref{Fig3}, $f_p$ vanishes at the axis of rotational symmetry and takes large values near the disk edges, leading to antisymmetric dip-peak patterns (\ref{Fig3}a). The resulting loss function $F_p$ (\ref{Fig3}b) exhibits oscillations depending on the relative phase with which the potential $f_p$ is sampled along the electron trajectory (see \ref{Fpxi}). As anticipated above, this depends on the path length traveled by the electron during an optical plasmon period ($v/\omega_p\sim v/\sqrt{E_F/D}$) relative to the extension of the plasmon ($\sim D$), the ratio of which is precisely $\mu/\nu$. When exciting symmetric plasmon modes ({\it e.g.}, $m=0$), non-oscillating $F_p$ profiles are obtained, equally characterized by maxima exceeding 1 at $\mu/\nu\sim2$ for electrons passing at a small distance from the disk (see Supplementary Information, SI).

For a relatively grazing trajectory ($\theta\sim80-85^\circ$), a maximum plasmon excitation probability $P_p^{\rm cla}\sim3\sqrt{\alpha\hbar c/E_FD}$ is reached for an electron velocity $v\sim(1/2)\sqrt{\alpha cE_FD}$ (see \ref{Fig3}b). As an indicative value, for a disk of diameter $D=50\,$nm doped to a Fermi energy $E_F=0.4$\,eV, similar to those fabricated in recent experiments \cite{paper212}, which can sustain $m=1$ plasmons of $0.21$\,eV energy, the maximum excitation probability is $P_p^{\rm cla}\sim0.8$ and is obtained using $\sim50$\,eV electrons. These magnitudes can be readily computed for other disk size and doping conditions {\it via} the scaling laws for the plasmon frequency $\omega_p\propto\sqrt{E_F/D}$ (\ref{wplast}), the maximum excitation probability $P_p^{\rm cla}\propto1/\sqrt{E_FD}$, and the electron energy $\propto E_FD$. Although arbitrarily large values of $P_p^{\rm cla}$ can be in principle achieved through reducing $E_F$ and $D$ (even while maintaining their ratio constant, and consequently, also $\omega_p$) the graphene size is limited to $D\sim10$\,nm, as plasmons in smaller islands are strongly quenched by nonlocal effects \cite{paper183}. With $D=10$\,nm and $E_F=0.4\,$eV, we have $0.48$\,eV plasmons that can be excited by 9\,eV electrons with $P_p^{\rm cla}=1.8$ probability per incident electron. This result is clearly outside the range of validity of linear response theory and clearly anticipates large multiple-plasmon excitation probabilities.

{\bf Quantum mechanical description.} The above classical formalism follows a long tradition of explaining electron energy-loss spectra within classical theory,\cite{paper149} under the assumption that the total excitation rate is small ({\it i.e.}, $P_p^{\rm cla}\ll1$), thus rendering multiple plasmon excitations highly unlikely. This is inapplicable to describe electron-driven plasmon generation in graphene, for which we can have $P_p^{\rm cla}>1$. Therefore, a quantum treatment of the plasmons becomes necessary. We follow a similar approach as in previous studies of multiple plasmon losses \cite{SL1971,MSG10,LMG13}, here adapted to deal with a single plasmon mode. Describing the electron as a classical external charge density $\rho^{\rm ext}(\rb,t)$ and the plasmon as a bosonic mode, we consider the Hamiltonian
\begin{equation}
H=\hbar\omega_pa^+a+g(t)\left(a^++a\right),
\label{H}
\end{equation}
where the operator $a$ ($a^+$) annihilates (creates) a plasmon of frequency $\omega_p$, and the time-dependent coupling coefficient is defined as
\begin{equation}
g(t)=\int d^3\rb\,\phi_p(\rb)\rho^{\rm ext}(\rb,t)
\label{gt}
\end{equation}
in terms of $\phi_p$, the electric potential associated with the plasmon. In the electrostatic approximation, neglecting the effect of inelastic plasmon decay, it is safe to assume that $\phi_p$ is real. Notice that $g$ is just the electrostatic energy subtracted or added to the system when removing or creating one plasmon ({\it i.e.}, the integral represents the potential energy of the external charge in the presence of the potential created by one plasmon). The Hamiltonian $H$ should be realistic under the condition that both the electron-plasmon interaction time and the optical cycle are small compared with the plasmon lifetime. In practice, this means that the electron behaves a point-like particle, or at least, its wave function is spread over a region of size $\ll v\tau$. Additionally, we assume the electron kinetic energy to be much larger than the plasmon energy, so that multiple plasmon excitations do not significantly change the electron velocity (non-recoil approximation).

As the plasmon state evolves under the influence of a linear term in \ref{H} with a classical coupling constant $g$, it should exhibit classical statistics \cite{CN1965}. Indeed, it is easy to verify that the plasmon wave function
\begin{equation}
|\psi\rangle=\ee^{\ii\chi(t)}|\xi(t)\rangle
\label{apsi}
\end{equation}
is a solution of Schr\"odinger's equation $H|\psi\rangle=\ii\partial|
\psi\rangle/\partial t$, where
\begin{equation}
|\xi(t)\rangle=\exp(-|\xi|^2/2)\,\sum_n\frac{(\xi a^+)^n}{n!} e^{-in\omega_pt} |0\rangle.
\label{psichi}
\end{equation}
is a coherent state \cite{G1963} with
\begin{equation}\xi(t)=\frac{-\ii}{\hbar}\int_{-\infty}^t dt'\,g(t')\,\ee^{\ii\omega_pt'}+\xi(-\infty),
\nonumber
\end{equation}
whereas
\[\chi(t)=\frac{-1}{\hbar}\int_{-\infty}^t dt'\,g(t')\,{\rm Re}\left\{\xi(t')\,\cos(\omega_pt')\right\}+\chi(-\infty)\]
is an overall phase that does not affect the plasmon-number distribution. The average number of plasmons excited at a given time is given by $|\xi(t)|^2$. We thus conclude that the probability of exciting $n$ plasmons simutaneoulsy follows a Poissonian distribution
\begin{equation}
P_p^{(n)}=\langle\psi|\left[(a^+)^na^n/n!\right]|\psi\rangle=\frac{|\xi|^{2n}}{n!}\ee^{-|\xi|^2},
\label{Ppn}
\end{equation}
which yields a second-order correlation $g^{(2)}(0)=1$.

Expressing $\xi$ and $g$ in terms of $\phi_p$, noticing that $\xi(-\infty)=0$ ({\it i.e.}, no plasmons present before interaction with the electron), and using a similar scaling as in the classical theory discussed above, we can write the probability of exciting $n$ plasmons as
\begin{equation}
P_p^{(n)}=\frac{\left(P_p^{\rm cla}\right)^n}{n!}\,\exp(-P_p^{\rm cla}),
\label{multiple}
\end{equation}
where $P_p^{\rm cla}=|\xi|^2=\mu^{-1}F_p(\mu/\nu)$ is the classical linear probability given by \ref{PF}, which coincides with the average number of excited plasmons per electron, $P_p^{\rm cla}=\sum_n nP_p^{(n)}$, and can take values above 1.

\begin{figure}
\begin{center}
\includegraphics[width=100mm,angle=0,clip]{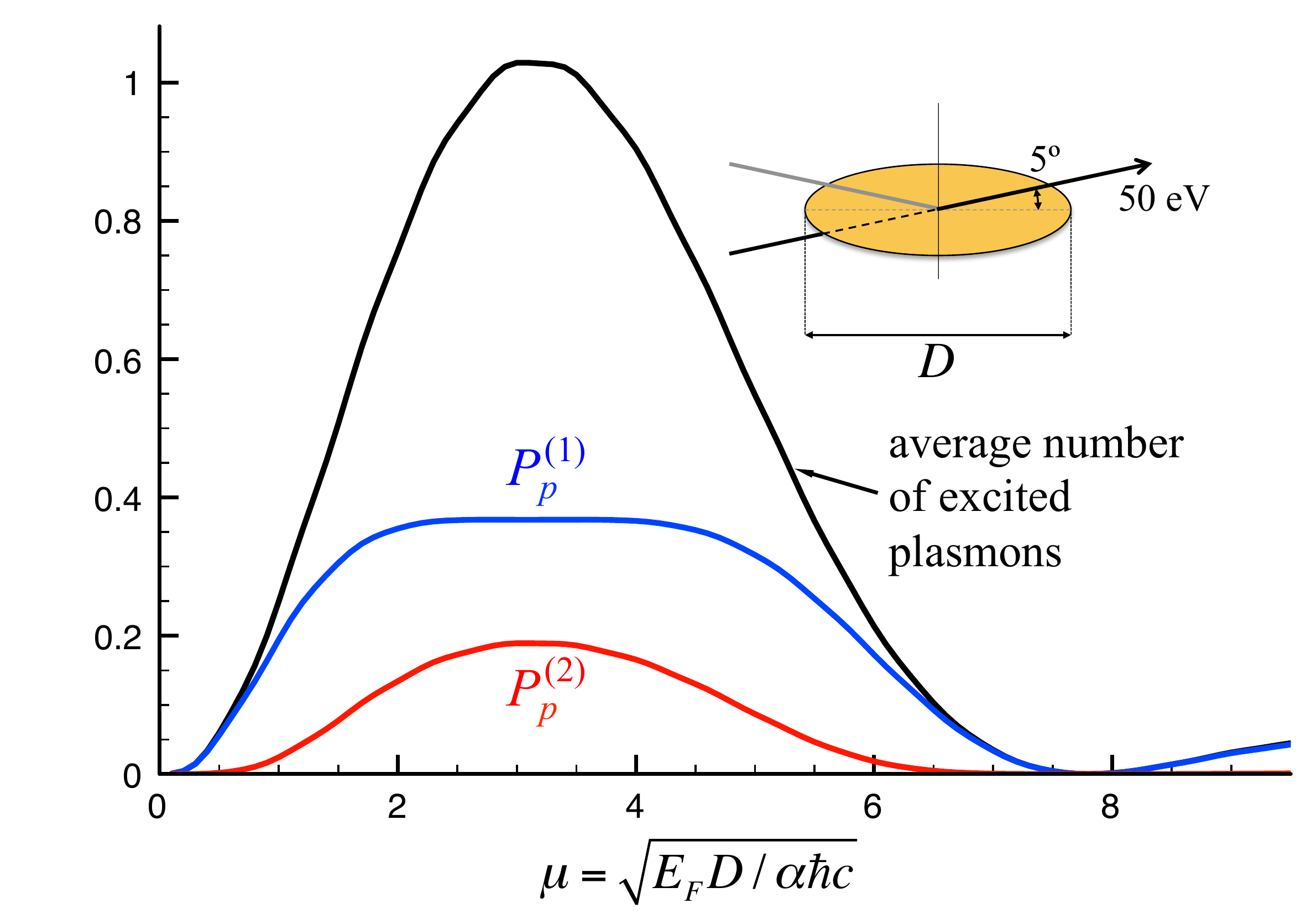}
\caption{{\bf Double plasmon excitation by a single electron.} The probabilities of exciting a single $m=1$ plasmon ($P_p^{(1)}$, blue curve) and two plasmons simultaneously ($P_p^{(2)}$, red curve) are compared with the average number of excited plasmons ($P_p^{\rm cla}=\sum_nnP_p^{(n)}$, black curve) for a 50\,eV electron passing grazingly by the center of a graphene disk. The probability is the same for straight crossing and specularly reflected trajectories.} \label{Fig4}
\end{center}
\end{figure}

{\bf Inclusion of plasmon losses.} Inelastic losses during the electron-plasmon interaction time have been so far ignored in the above quantum description. However, for sufficiently slow electrons or very lossy plasmons, the interaction time can be comparable to the plasmon lifetime $\tau$, so that the above quantum formalism needs to be amended, for example by following the time evolution of the density matrix $\rho$, according to its equation of motion \cite{FT02}
\begin{equation}
\frac{d\rho}{dt}=\frac{\ii}{\hbar}[\rho,H]+\frac{1}{2\tau}\left(2a\rho\,a^+-a^+a\;\rho-\rho\;a^+a\right),
\label{drhodt}
\end{equation}
where the Hamiltonian $H$ is defined by \ref{H,gt}. The solution to this equation can still be given in analytical form:
\begin{equation}
\rho=|\xi\rangle\langle\xi|,
\label{xixi}
\end{equation}
where $|\xi\rangle$ is again a coherent state (see \ref{psichi}), but we have to redefine
\begin{equation}
\xi(t)=\frac{-\ii}{\hbar}\int_{-\infty}^t dt'\,g(t')\,\ee^{\ii\omega_pt'-(t-t')/2\tau}+\xi(-\infty).
\label{newxi}
\end{equation}
It is straightforward to verify that \ref{xixi,newxi} are indeed a solution of 
\ref{drhodt}.

A simultaneous density-matrix description of the electron and plasmon quantum evolutions involves a larger configuration space that is beyond the scope of this paper. We can however argue that the probabilities $P_p^{(n)}$ corresponding to the electron lossing energies $n\hbar\omega_p$ must still follow a Poissonian distribution if we trace out the plasmon mode. At $t\rightarrow\infty$, all plasmons must have decayed, so that we can obtain the average number of plasmon losses from the time integral of the total plasmon decay rate, $\langle a^+a\rangle/\tau$. Again, we find that this quantity coincides with the classical linear loss probability $P_p^{\rm cla}$, which is now given by
\begin{equation}
P_p^{\rm cla}=\frac{1}{\tau}\int dt\;|\xi(t)|^2=\frac{1}{\hbar^2\tau}\int\frac{d\omega}{2\pi}\frac{|\tilde{g}(\omega)|^2}{(\omega_p-\omega)^2+1/4\tau^2},
\nonumber
\end{equation}
where $\tilde{g}(\omega)$ is the time-Fourier transform of $g(t)$. This expression reduces to $P_p^{\rm cla}=|\tilde{g}(\omega_p)/\hbar|^2$ ({\it i.e.}, \ref{PF}) in the limit of high plasmon quality factor $Q=\omega_p\tau\gg1$.

{\bf Multiple plasmon generation by a single electron.} We show in \ref{Fig4} results obtained by solving \ref{multiple} under the conditions of the most grazing trajectory from those considered in \ref{Fig3}. In particular, the electron energy is 50\,eV, which corresponds to $\nu\approx1.9$. The average number of plasmons excited by a single electron under these conditions reaches a maximum value slightly above 1, distributed in a $\sim40\%$ probability of exciting only one plasmon, a $\sim20\%$ probability of simultaneously exciting two plasmons, and lower probabilities of generating more than two plasmons. The peak maximum is observed at $\mu\sim3$ (this corresponds for example to $D=50$\,nm and $E_F=0.26$\,eV), which leads to $\mu/\nu\sim1.6$, slightly to the left of the main peak observed within linear theory at $\mu/\nu\sim2$ in \ref{Fig3}.

\begin{figure}
\begin{center}
\includegraphics[width=160mm,angle=0,clip]{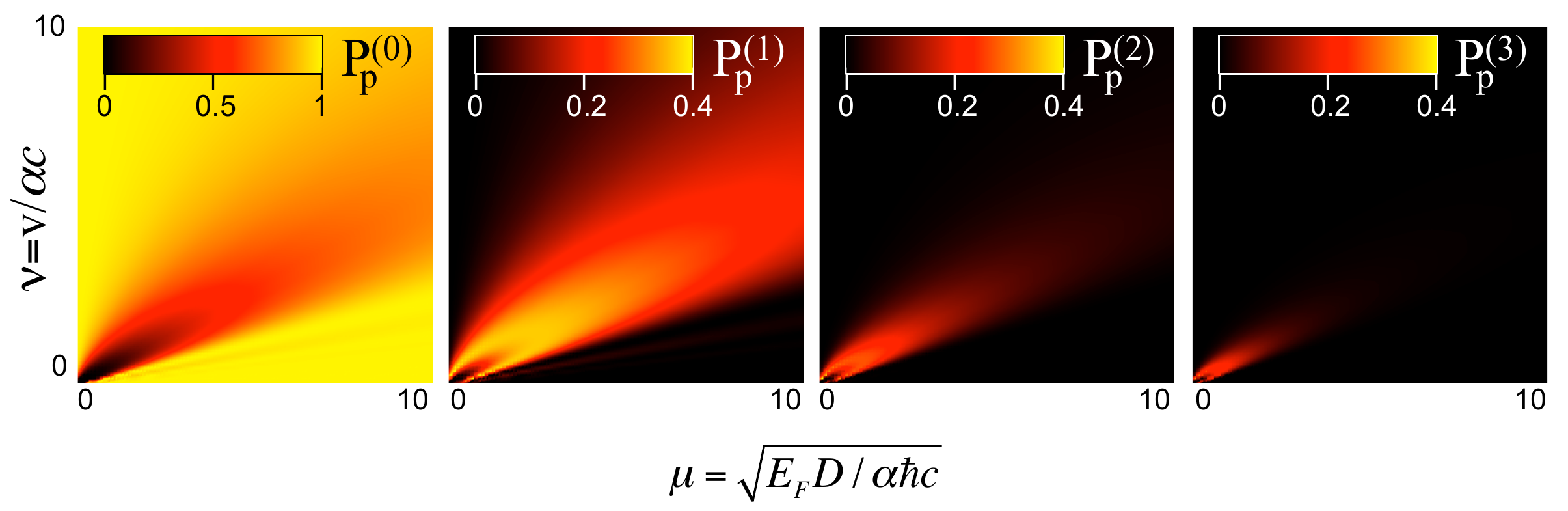}
\caption{{\bf Multiple-plasmon excitation by a single electron.} Full $\mu$ and $\nu$ dependence of the probabilities $P_p^{(n)}$ of generating $n=0-3$ plasmons under the same conditions as in \ref{Fig4}.} \label{Fig5}
\end{center}
\end{figure}

The full dependence of the probability of generating $n=0-3$ plasmons on $\mu$ and $\nu$ is shown in \ref{Fig5}. For large $\mu$, the results approach the linear regime, only single plasmons are effectively excited, and the highest probability is peaked around a broad region centered along the $\mu=2\,\nu$ line. At small $\mu$'s, a more complex behavior is observed. The double-plasmon excitation probability is above 20\% over a broad range of $\mu$'s, and even the probability of simultaneously generating three plasmons takes significant values $>10\%$ up to $\mu\sim1$. The probability is increasingly more confined towards the low $\mu$ and $\nu$ region when a larger number of plasmons is considered. Notice however the presence of a dip in that region for single-plasmon excitation, which is due to transits towards a larger numbers of plasmons created. A similar effect is observed for $n=2$ at even lower values of $\mu$ and $\nu$.

\section{CONCLUSIONS AND OUTLOOK}

We predict unprecedentedly high graphene-plasmon excitation rates by relatively low-energy free electrons. When a plasmon is highly confined down to a small size compared with the light wavelength, the dipole moment associated with the plasmon is small and this limits the strength of its coupling to light. Direct optical excitation becomes inefficient, and one requires near-field probes to couple to the plasmons \cite{paper196,FRA12}. In fact electrons act as versatile near-field probes that can be aimed at the desired sample region. Furthermore, electrons carry strongly evanescent electromagnetic fields that couple with high efficiency to confined optical modes, thus rendering the observation of multiple plasmon excitation feasible. Our calculations show double excitation of a single plasmon mode with efficiencies of up to 20\% per incident electron in graphene structures with similar size and doping levels as those produced in recent experiments \cite{paper212}.

Besides its fundamental interest, multiple plasmon excitation is potentially useful to explore nonlinear optical response at the nanoscale. In particular, quantum nonlinearities at the single- or few-plasmon level has been predicted in small graphene islands \cite{GCK14}. As the excitation of strongly confined plasmons by optical means remains a major challenge, free electrons provide the practical means to explore this exotic quantum behavior. In particular, the generation of $n$ plasmons by a single electron produces an energy loss $\hbar n\omega_p$, the detection of which can be used to signal the creation of a plasmon-number state in the graphene island.

The electron energies for which plasmon excitation probabilities are high lie in the sub-keV range, which is routinely employed in low-energy EELS studies \cite{R95}, as already reported for collective excitations in graphene \cite{LWE08,LW10}. In practice, one could study electrons that are elastically (and specularly) reflected on a patterned graphene film. Actually, the analysis of energy losses in reflected electrons was already pioneered in the first observation of surface plasmons in metals \cite{PS1959}. A similar approach could be followed to reveal multiple graphene plasmon excitations by individual electrons. Incidentally, low-energy electron diffraction at the carbon honeycomb lattice could provide additional ways of observing these excitations through elastically diffracted beams, for which plasmons should be the dominant channel of inelastic losses. Inelastic losses in core-level photoelectrons, which have been used to study plasmons in semi-infinite \cite{OGH1990,AHK96} and ultrahin \cite{OME11} metals, as well as  (the so-called plasmon satellites), offer another alternative to resolve multiple plasmon excitations in confined systems.

In practical experiments, the graphene structures are likely to lie on a substrate, and their size and shape must be chosen such that the energies of higher-order plasmons are not multiples of the targeted plasmon energy. Although for the sake of simplicity the analysis carried out here is limited to self-standing graphene structures, it can be trivially extended to carbon islands supported on a substrate of permittivity $\epsilon$ by simply multiplying both the graphene conductivity (or equivalently, the Fermi energy in the Drude model) and the external electron potential by a factor $2/(1+\epsilon)$. This factor incorporates a rigorous correction to the $1/r$ free-space point-charge Coulomb interaction when the charge is instead placed right at the substrate surface. Likewise, the contribution of the image potential leads to a total external potential at the surface given by $2/(1+\epsilon)$ times the bare potential of the moving electron. The presence of a surface can however attenuate the transmitted electron intensity, so reflection measurements appear as a more suitable configuration. It should be emphasized that the energy loss and plasmon excitation probabilities produced by transmitted electrons coincide with those for specularly reflected electrons, and therefore, the present theory is equally applicable to reflection geometries, as indicated in \ref{Fig4}. This is due to the small thickness of the graphene, where induced charges cannot provide information on which side the electron is coming from. A detailed analysis consisting in using an external electron charge $\rho^{\rm ext}(\rb,t)$ for a specularly reflected electron fully confirms this result.

High plasmon excitation efficiencies should be also observable in metal nanoparticles. One could for example study electrons reflected on a monolayer of nanometer-sized gold colloids, which present a similar degree of mode confinement as graphene. However, plasmons in noble metals have lower optical quality factors than in graphene, thus compromising the condition that $v\tau$ be smaller than the electron wave function spread (see above). Additionally, the trajectories of sub-keV electrons reflected from metal colloids can be dramatically affected by their stochastic distributions of facets and small degree of surface homogeneity compared with graphene, the 2D morphology of which can be tailored with nearly atomic detail \cite{BFG12}.

In summary, graphene provides a unique combination of surface quality, tunability, and optical confinement that makes the detection of multiple plasmon excitations by individual electrons feasible, thus opening a new avenue to explore fundamental quantum phenonema, nanoscale optical nonlinearities, and efficient mechanisms of plasmon excitation and detection with potential application to opto-electronic nanodevices. 

\section{METHODS}

In this section, we formulate an electrostatic scaling law and a quantum-mechanical model that allow us to describe multiple excitations of graphene plasmons by fast electrons. The model agrees with classical theory within first-order perturbation theory, and provides a fast, accurate procedure to compute excitation probabilities for a wide range of electron velocities and graphene parameters.

{\bf Eigenmode expansion of the classical electrostatic potential near graphene.} The graphene structures under consideration are much smaller than the light wavelengths associated with their plasmon frequencies, and therefore, we can describe their response in the electrostatic limit. The optical electric field can be thus expressed as the gradient of an electric potential $\phi$ in the plane of the graphene. It is convenient to write the self-consistent relation
\begin{equation}
\phi(\rb,\omega)=\phi^{\rm ext}(\rb,\omega)+\frac{\ii}{\omega}
\int\frac{d^2\Rb'}{|\rb-\Rb'|}\nabla_{\Rb'}\cdot\sigma(\Rb',\omega)\nabla_{\Rb'}\phi(\Rb',\omega),
\label{eq1}
\end{equation}
where the integral represents the potential produced by the charge density induced on the graphene, which in virtue of the continuity equation, is in turn expressed as $(-\ii/\omega){\nabla_\Rb}\cdot\jb(\Rb,\omega)$ in terms of the induced current $\jb(\Rb,\omega)=-\sigma(\Rb,\omega){\nabla_\Rb}\phi(\Rb,\omega)$. Here, $\sigma$ is the 2D graphene conductivity, which we assume to act locally. These expressions involve coordinate vectors $\Rb=(x,y)$ in the plane of the graphene, $z=0$. Although \ref{eq1} is valid for any point $\rb=(\Rb,z)$, we take $z=0$ to obtain the self-consistent electric potential in the graphene sheet. Incidentally, the abrupt change of $\sigma$ at the edge of a graphene structure produces a divergent boundary contribution to the integrand of \ref{eq1}. These types of divergences have been extensively studied in the context of magneplasmons at the edge of a bounded 2D electron gas \cite{F1986}, and more recently also in graphene \cite{WKA12}. In practice, we can solve \ref{eq1} numerically by smoothing the edge (e.g., though in-plane modal expansions of $\sigma$ and $\phi$). This produces convergent results in agreement with direct solutions of the 3D Poisson equation. However, we only use \ref{eq1} in this article to derive formal relations and scaling laws involving plasmon modes, whereas specific numerical computations are performed following a different method, as explained below.

Given the lack of absolute length scales in electrostatics, we can recast \ref{eq1} in scale-invariant form by using the reduced 2D coordinate vectors $\th=\Rb/D$, where $D$ is a characteristic length of the graphene structure ({\it e.g.}, the diameter of a disk). Additionally, we assume that the conductivity can be separated as $\sigma(\Rb,\omega)=f(\Rb)\sigma(\omega)$. For homogeneously doped graphene, $f(\Rb)$ simply represents a filling factor that takes the value $f=1$ in the graphene and vanishes elsewhere. However, the present formalism can be readily applied to more realistic inhomogeneous doping profiles by transferring the space-dependence of $E_F$ to $f$. This can be applied to describe inhomogeneously doped graphene, including divergences in the doping density near the edges \cite{paper194}. Combining these elements, we obtain
\begin{equation}
\phi(\th,\omega)=\phi^{\rm ext}(\th,\omega)+\eta(\omega)
\int\frac{d^2\th'}{|\th-\th'|}\nabla_{\th'}\cdot f(\th')\nabla_{\th'}\phi(\th',\omega),
\label{eq2}
\end{equation}
where
\begin{equation}
\eta(\omega)=\frac{i\sigma(\omega)}{\omega D}
\label{eta}
\end{equation}
is a dimensionless parameter containing all the physical characteristics of the graphene, such as the doping level, the temperature dependence, and the rate of inelastic losses, as well as the dependence on frequency $\omega$. Integrating by parts and taking the in-plane 2D gradient on both sides of \ref{eq2}, we find the more symmetric expression
\begin{equation}
\fE(\th,\omega)=\fE^{\rm ext}(\th,\omega)+\eta(\omega)
\int d^2\th'\;\Mb(\th,\th')\cdot\fE(\th',\omega),
\label{fEfE}
\end{equation}
where $\fE(\th,\omega)=-\sqrt{f(\th)}\,{\nabla_{\th}}\phi(\th,\omega)$ and
\begin{equation}
\Mb(\th,\th')=\sqrt{f(\th)f(\th')}\;\;{\nabla_{\th}}\otimes{\nabla_{\th}}\,\frac{1}{|\th-\th'|}
\nonumber
\end{equation}
is a symmetric matrix that is invariant under exchange of its arguments: $\Mb(\th,\th')=\Mb(\th',\th)$. This implies that $M$ is a real, symmetric operator that admits a complete set of real eigenvalues $1/\eta_j$ and orthonormalized eigenvectors $\fE_j$ satisfying
\begin{equation}
\fE_j(\th)=\eta_j\int d^2\th'\;\Mb(\th,\th')\cdot\fE_j(\th'),
\nonumber
\end{equation}
\begin{equation}
\int d^2\th\;\;\fE_j(\th)\cdot\fE_{j'}(\th)=\delta_{jj'},
\nonumber
\end{equation}
and
\begin{equation}
\sum_j\fE_j(\th)\otimes\fE_j(\th')=\delta(\th-\th')\;\mathbb{I}_2,
\nonumber
\end{equation}
where $\mathbb{I}_2$ is the $2\times2$ unit matrix.

The solution to \ref{fEfE} can be expressed in terms of these eigenmodes as
\begin{equation}
\fE(\th,\omega)=\sum_j\frac{c_j}{1-\eta(\omega)/\eta_j}\,\fE_j(\th),
\nonumber
\end{equation}
with expansion coefficients
\begin{equation}
c_j=\int d^2\th\;\;\fE_j(\th)\cdot\fE^{\rm ext}(\th,\omega).
\label{cj}
\end{equation}
The potential outside the graphene can be constructed from $\phi(\th,\omega)$ through the induced charge $(i\sigma(\omega)/\omega){\nabla_\Rb}\cdot f(\Rb){\nabla_\Rb}\phi(\Rb,\omega)=(-\eta/D)\,\nabla_{\th}\cdot\sqrt{f(\th)}\fE(\th,\omega)$. Following a procedure similar to the derivation of \ref{eq2}, we find
\begin{equation}
\phi^{\rm ind}(\ub,\omega)=\sum_j\frac{c_j}{1/\eta_j-1/\eta(\omega)}\,\varphi_j(\ub),
\label{phiind}
\end{equation}
where
\begin{equation}
\varphi_j(\ub)=\int\frac{d^2\th'}{|\ub-\theta'|}\;\nabla_{\th'}\cdot\sqrt{f(\th')}\;\fE_j(\th')
\label{varphij}
\end{equation}
and we have defined the reduced 3D coordinate vector $\ub=\rb/D$.

{\bf Classical screened interaction potential.} The screened interaction $W^{\rm ind}(\rb,\rb',\omega)$ is defined as the potential produced at $\rb$ by a unit point charge placed at $\rb'$ and oscillating in time as $\exp(-\ii\omega t)$. We use the formalism introduced in the previous paragraph and consider in \ref{cj} the point-charge external potential $\phi^{\rm ext}(\th,\omega)=(1/D)/|\th-\ub'|$. Integrating by parts, we find $c_j=\varphi_j(\ub')/D$ (see \ref{varphij}), which upon insertion into \ref{phiind} yields
\begin{equation}
W^{\rm ind}(\ub,\ub',\omega)=\frac{1}{D}\sum_j\frac{1}{1/\eta_j-1/\eta(\omega)}\,\varphi_j(\ub)\varphi_j(\ub').
\label{W}
\end{equation}
The well-known symmetry $W^{\rm ind}(\ub,\ub',\omega)=W^{\rm ind}(\ub',\ub,\omega)$ is apparent in \ref{W}. For an arbitrary external charge distribution $\rho^{\rm ext}(\rb,t)$, which we express in frequency space as
\begin{equation}
\rho^{\rm ext}(\rb,\omega)=\int dt\,\ee^{\ii\omega t}\rho^{\rm ext}(\rb,t),
\label{rhow}
\end{equation}
the induced potential can be written using the screened interaction as
\begin{equation}
\phi^{\rm ind}(\rb,\omega)=\int d^3\rb'\;W^{\rm ind}(\rb,\rb',\omega)\,\rho^{\rm ext}(\rb',\omega).
\label{Wrho}
\end{equation}

\begin{figure}
\begin{center}
\includegraphics[width=60mm,angle=0,clip]{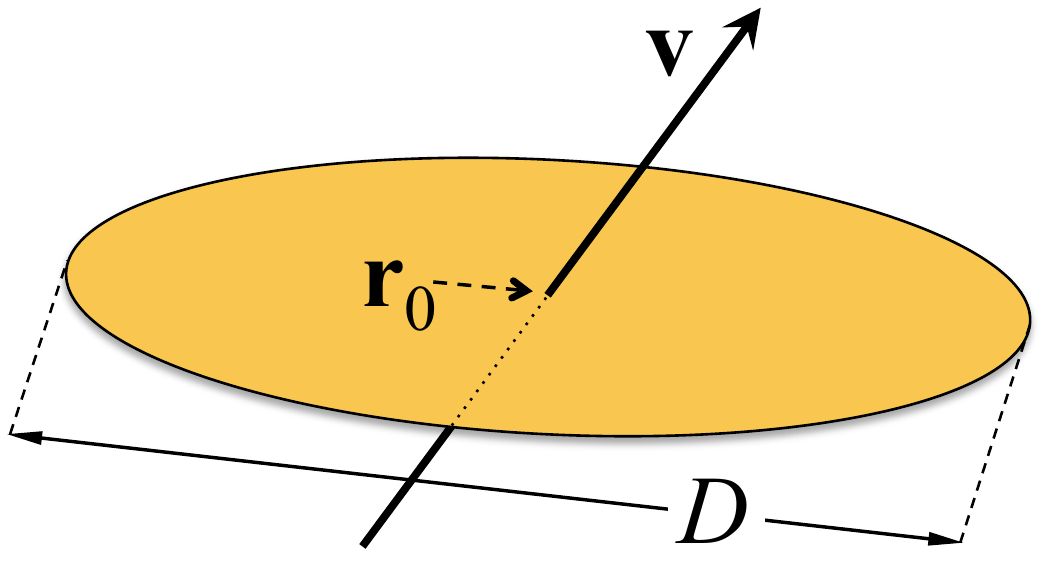}
\caption{Electron moving with velocity $\vb$ and crossing a graphene structure of characteristic size $D$.}
\label{FigS1}
\end{center}
\end{figure}

{\bf Electrostatic scaling law for the plasmon frequency.} The condition $\eta(\omega_j)=\eta_j$ determines a plasmon frequency of the system $\omega_j$ for a specific eigenstate $j$, subject to the condition $\eta_j<0$ (see below). This relation has general validity within the electrostatic limit, so that all the graphene characteristics, including the size of the structure $D$, are fully contained within $\eta(\omega)$, and thus, given a certain shape ({\it e.g.}, a disk), the eigenvalues $\eta_j$ can be calculated once and for all to obtain the plasmon frequency for arbitrary size or doping.

A powerful electrostatic law can be formulated by assuming the Drude model for the conductivity of graphene (\ref{Drude}). The plasmon frequency is then given by $\omega_j-\ii/2\tau$, where
\begin{equation}
\omega_j=\sqrt{\frac{e^2E_F}{-\pi\eta_j\hbar^2D}-\frac{1}{4\tau^2}}\approx\gamma_j\frac{e}{\hbar}\sqrt{\frac{E_F}{D}}
\label{wp}
\end{equation}
and we have defined the real number (provided $\eta_j<0$)
\begin{equation}
\gamma_j=1/\sqrt{-\pi\eta_j}
\label{gamp}
\end{equation}
to obtain the rightmost expression in \ref{wp}.

{\bf Classical approach to the electron energy-loss probability.} We consider an electron moving with constant velocity vector $\vb$ along the straight-line trajectory $\rb=\rb_0+\vb t$ passing near or through a graphene structure, as shown in \ref{FigS1}. The energy transferred from the electron to the graphene ($\Delta E>0$) can be written as \cite{paper149}
\begin{equation}
\Delta E=\int d\omega\,\hbar\omega\,\Gamma(\omega),
\nonumber
\end{equation}
where
\begin{equation}
\Gamma(\omega)=\frac{e}{\pi\hbar}\int dt\;{\rm Im}\left\{\ee^{-\ii\omega t}\;\phi^{\rm ind}(\rb_0+\vb t,\omega)\right\}
\label{P}
\end{equation}
is the loss probability per unit of frequency range, and $\phi^{\rm ind}$ is the $\omega$ component of the potential induced by the electron along its path. The external charge density associated with the moving electron reduces to $\rho^{\rm ext}(\rb,t)=-e\,\delta(\rb-\rb_0-\vb t)$. Using this expression together with \ref{W,rhow,Wrho,P}, the loss probability is found to be
\begin{equation}
\Gamma(\omega)=\frac{e^2}{\pi\hbar\omega^2D}\,\sum_j{\rm Im}\left\{\frac{1}{1/\eta(\omega)-1/\eta_j}\right\}\;G_j(\zeta),
\label{PG}
\end{equation}
where $\zeta=\omega D/v$,
\begin{equation}
G_j(\zeta)=\left|\zeta\int d\ell\;\ee^{\ii\zeta	\ell}\;\varphi_j(\ub_0+\ell\hat{\vb})\right|^2,
\label{Gj}
\end{equation}
we adopt the notation $\ub=\rb/D$, and the integral is over the path length $\ell$ (in units of $D$) along the velocity vector direction.

We now concentrate on a specific plasmon resonance $j$ and neglect contributions to the loss probability arising from modes other than this particular one. For simplicity, we work within the Drude model and assume a small plasmon width $\tau^{-1}\ll\omega_j$. Using \ref{eta,Drude} in \ref{PG}, and integrating over $\omega$ to cover the plasmon peak area, we find
\begin{equation}
P_j^{\rm cla}=\int_j d\omega\;\Gamma(\omega)\approx\frac{1}{2\pi^2\gamma_j^3}\frac{e}{\sqrt{E_FD}}\;\;G_j\!\left(\gamma_j\frac{e\sqrt{E_FD}}{\hbar v}\right)
\label{Pj}
\end{equation}
for the probability of exciting plasmon $j$ by the incident electron. Finally, we can recast \ref{Pj} into the scale-invariant form of \ref{PF}, using the definitions of \ref{munu,Fpxi}, and further defining the dimensionless scaled plasmon electric potential
\begin{equation}
f_j=\frac{1}{\sqrt{2\pi\gamma_j}}\varphi_j.
\label{fj}
\end{equation}

\begin{figure}
\begin{center}
\includegraphics[width=100mm,angle=0,clip]{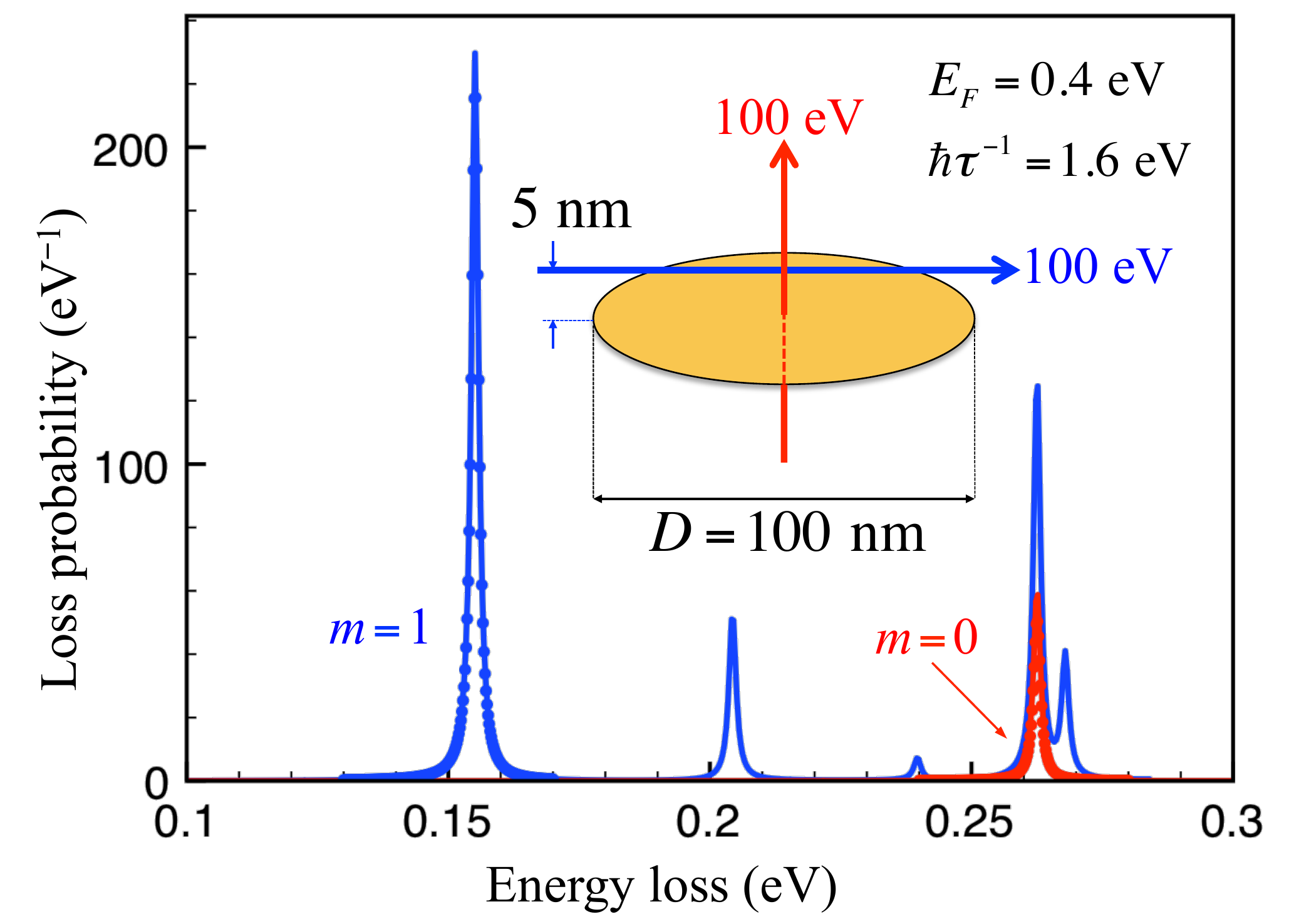}
\caption{Energy-loss spectra for 100\,eV electrons passing near a graphene disk along the trajectories shown in the inset. The graphene parameters are indicated by labels. Solid curves are calculated with the boundary-element method \cite{paper040} using the local-RPA model for the graphene conductivity \cite{FV07}. The results obtained from the semi-analytical model of \ref{PG} are represented by symbols for the lowest-order $m=0$ and $m=1$ modes, in excellent agreement with the solid curves (see also \ref{Fig2}).}
\label{FigS2}
\end{center}
\end{figure}

We show in \ref{FigS2} two examples of loss spectra for 100\,eV electrons passing near a 100\,nm graphene disk. The lowest-energy plasmon features for both $m=0$ and $m=1$ symmetries are clearly separated from other peaks in their corresponding spectral regions, thus justifying the approximation of \ref{Pj}. We further compare in this figure the full solution of Maxwell's equations (solid curves) with the result obtained from \ref{PG} using a single plasmon term for each of the lowest-order $m=0$ and $m=1$ modes, with $\eta_j=-0.024$ and $-0.073$, respectively, and with $\varphi_j$ calculated from the plasmon potential, which is normalized as explained below.

{\bf Quantum approach to the screened interaction potential.}  The linear screened interaction potential can be obtained by solving the density matrix (\ref{drhodt}), which yields the solution $\rho=|\xi\rangle\langle\xi|$, with $|\xi\rangle$ given by \ref{psichi,newxi}. Expressing $g(t)$ in frequency space $\omega$, we can then write
\begin{equation}
\xi(t)=\frac{-1}{\hbar} \int d^3\rb \int\frac{d\omega}{2\pi}\;\phi_j(\rb)\,\rho^{\rm ext}(\rb,\omega)\;\frac{\ee^{\ii(\omega_j-\omega)t}}{\omega_j-\omega-\ii/2\tau},
\nonumber
\end{equation}
where $\rho^{\rm ext}(\rb,\omega)$ is defined by \ref{rhow}. Now, calculating the induced potential from its expectation value
\begin{equation}
\phi^{\rm ind}(\rb,t)=\langle\xi|\left(a^++a\right)\phi_j(\rb)|\xi\rangle,
\nonumber
\end{equation}
we find
\begin{equation}
\phi^{\rm ind}(\rb,t)=\int\frac{d\omega}{2\pi} \ee^{-\ii\omega t} \int d^3\rb' W^{\rm ind}(\rb,\rb',\omega)\rho^{\rm ext}(\rb',\omega)
\nonumber
\end{equation}
(or equivalently, \ref{Wrho}), where
\begin{equation}
W^{\rm ind}(\rb,\rb',\omega)= \frac{2\omega_j}{\hbar}\frac{\phi_j(\rb)\phi_j(\rb')}{(\omega+\ii/2\tau)^2-\omega_j^2}
\label{WQM}
\end{equation}
is the quantum-mechanical counterpart of \ref{W}.

{\bf Normalization of the plasmon potential.} In the Drude model (\ref{Drude}), assuming a dominant plasmon mode contributing to the response with frequency given by \ref{wplast}, we find that \ref{W,WQM} are identical under the assumption $\omega_j\tau\gg1$, provided we take
\begin{equation}
\phi_j=\left(\frac{e^2E_F}{D^3}\right)^{1/4}\;f_j,
\label{phivarphi}
\end{equation}
where $f_j$ is the dimensionless scaled plasmon electric potential defined by \ref{fj}, which is independent of the doping level $E_F$ and the size of the structure $D$. As a self-consistency test, we calculate the plasmon excitation probability to first-order perturbation from the quantum model, which yields
\begin{equation}
P_j=\frac{e^2}{\hbar^2}\left|\int dt\;\ee^{\ii\omega_jt}\;\phi_j(\rb_0+\vb t)\right|^2.
\label{PpQ}
\end{equation}
Indeed, this equation coincides with the classsical result of \ref{Pj}, provided \ref{phivarphi} is satisfied.

In practice we obtain the plasmon potential $\phi_j$ as follows: first, we calculate the potential induced by a dipole placed near the graphene and oscillating at frequency $\omega_j$ using the boundary-element method \cite{paper040} (BEM) for fixed values of $E_F$ and $D$; the resulting potential must be equal to $\phi_j$ times an unknown constant; we deduce this constant by calculating the plasmon excitation probability from \ref{PpQ} and by comparing the result to a well-established classical calculation of the loss probability based upon BEM \cite{paper149}; finally, we use the scaling laws of \ref{wp,phivarphi} to obtain $\omega_j$ and $\phi_j$ for any desired values of $E_F$ and $D$, assuming the validity of the Drude model. We have verified that this procedure yields, within the accuracy of the BEM method, the same scaled potentials $f_j$ and $\varphi_j$ for different initial values of $E_F$ and $D$. Incidentally, once $\phi_j$ is calculated, \ref{PpQ} provides a fast way of obtaining loss probabilities for arbitrary values of the electron velocity, the size of the structure, and the doping conditions.

{\bf Analytical expressions for the electron energy-loss probability in homogeneous graphene.} In the electrostatic limit, the loss probability of electrons moving either parallel or perpendicularly with respect to an extended sheet of homogeneously doped graphene can be expressed in terms of the Fresnel reflection coefficient for $p$ polarized light, as shown in \ref{Gperp,Gpara}. Indeed, for a parallel trajectory, we can readily use the well-established dielectric formalism ({\it e.g.}, Eq.\ (25) of Ref.\ \cite{paper149} in the $c\rightarrow\infty$ limit), which directly yields \ref{Gpara}. Likewise, for a perpendicular trajectory, we can write the bare potential of the moving electron in $(\kb_\parallel,\omega)$ space as
\begin{equation}
\phi^{\rm ext}(k_\parallel,z)=\frac{-4\pi e}{v}\frac{\ee^{\ii\omega z/v}}{k_\parallel^2+\omega^2/v^2}.
\nonumber
\end{equation}
Inserting this expression into \ref{eq1} and writing the 2D Coulomb interaction as $(2\pi/k_\parallel)\ee^{-k_\parallel|z|}$, we find the induced potential
\begin{equation}
\phi^{\rm ind}(k_\parallel,z)=-r_p\,\ee^{-k_\parallel|z|}\,\phi^{\rm ind}(k_\parallel,0),
\nonumber
\end{equation}
which, together with \ref{P}, allows us to write the loss probability for a perpendicular trajectory as shown in \ref{Gperp}. It should be noted that the external field produced on the graphene by an electron that is specularly reflected at the graphene plane is also given by the above expression for $\phi^{\rm ind}(k_\parallel,0)$, and consequently, \ref{Gperp} yields the loss probability for such reflected trajectory as well.

\section{Supporting Information}

We provide further examples of the normalized functions $f_j$ and $F_j$, as well as multiple plasmon excitations for parallel and perpendicular trajectories with respect to a doped graphene disk.

\begin{figure}
\begin{center}
\includegraphics[width=170mm,angle=0,clip]{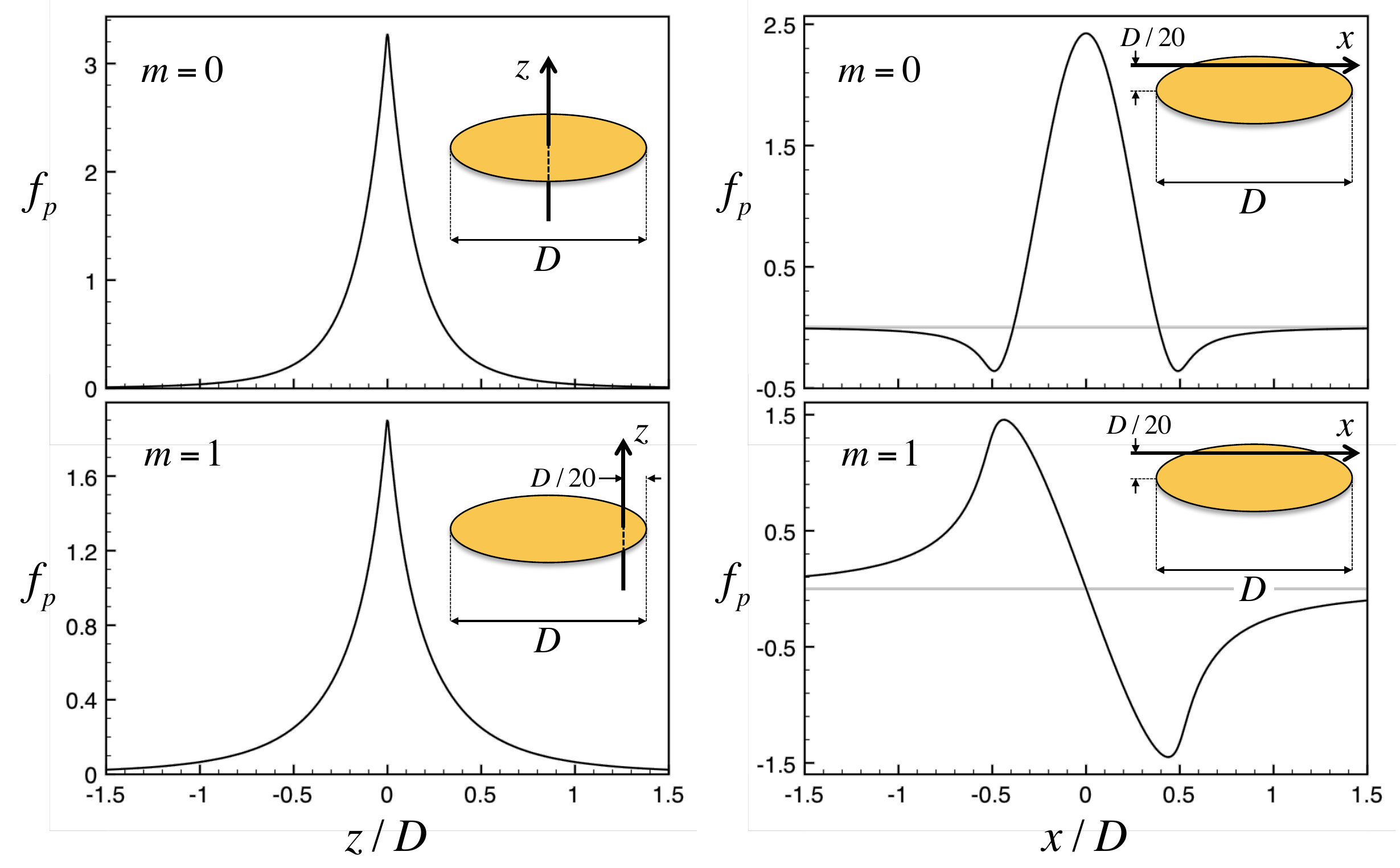}
\caption{Normalized electrostatic potential $f_j$ for the lowest-order $m=0$ and $m=1$ modes of a homogeneously doped graphene disk and for different sampling trajectories. The $m=0$ and $m=1$ lowest-order modes have energies given by $\hbar\omega_j=\gamma_je\sqrt{E_F/D}$ with $\gamma_j=3.6$ and $\gamma_j=2.0$, respectively. The parallel electron trajectory in the right panels crosses the disk axis at $x=0$.}
\label{FigS3}
\end{center}
\end{figure}

\begin{figure}
\begin{center}
\includegraphics[width=140mm,angle=0,clip]{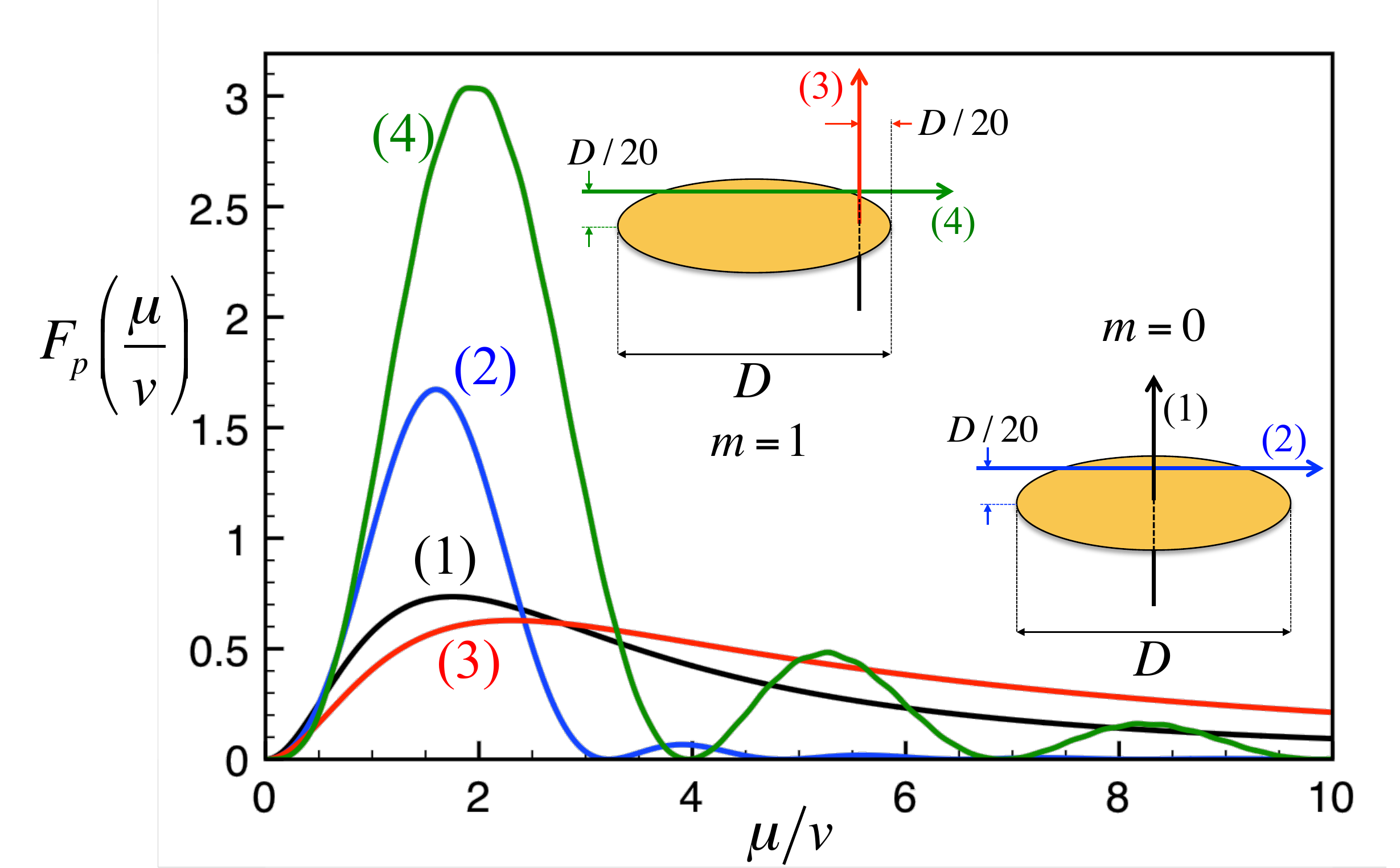}
\caption{Normalized loss function $F_j$ for the lowest-order $m=0$ and $m=1$ modes of a homogeneously doped graphene disk and for different sampling trajectories.}
\label{FigS4}
\end{center}
\end{figure}

Examples of the normalized functions $f_j$ and $F_j$ for a disk and different electron trajectories are offered in Figs.\ \ref{FigS3} and \ref{FigS4}. The highest probability among the trajectories considered in these figures is $F_j\approx3.0$, which is obtained in the excitation of the $m=1$ mode with $\mu/\nu\approx1.9$ for an electron moving parallel to the graphene surface. With a characteristic doping $E_F=0.5\,$eV and a disk diameter $D=50\,$nm, we have $\mu\approx4.2$, and the excitation peak occurs for $\nu=2.2$, or equivalently, for an electron energy $\approx65\,$eV. The plasmon energy $\hbar\omega_j=\gamma_je\sqrt{E_F/D}$ (see main paper) is $0.24\,$eV. The electron passes at a distance $D/20=5\,$nm from the graphene (see Fig.\ \ref{FigS4}, upper inset). The excitation probability $P_j\approx(1/\mu)\;F_j(\mu/\nu)$ is then 73\%, which reveals a clear departure from the perturbative regime, and thus this configuration requires a more detailed analysis including multiple-plasmon excitation.

\begin{figure}
\begin{center}
\includegraphics[width=160mm,angle=0,clip]{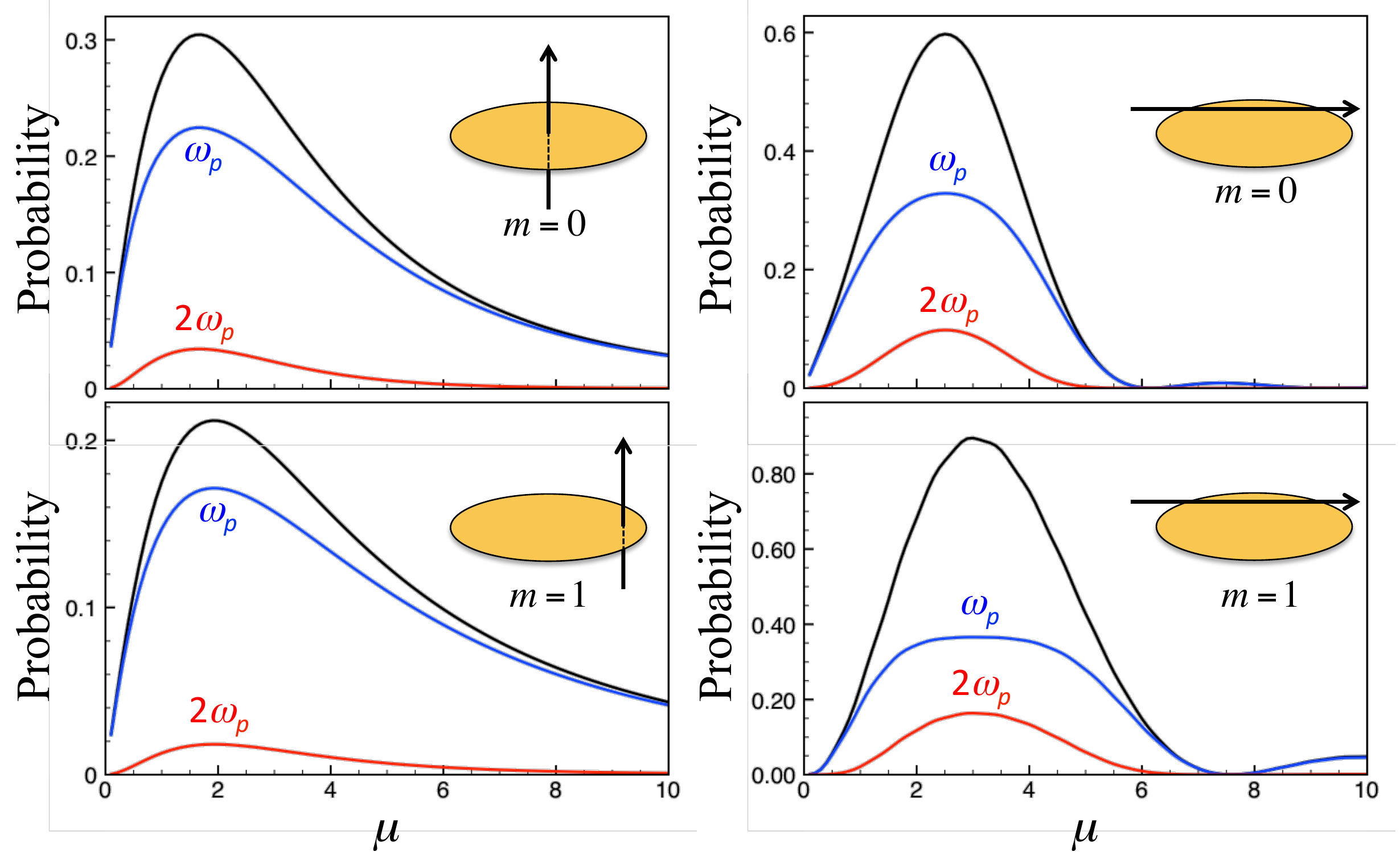}
\caption{Probability of exciting the lowest-order $m=0$ and $m=1$ modes of a homogeneously doped graphene disk for the same electron trajectories as considered in Figs.\ \ref{FigS3} and \ref{FigS4}, as a function of $\mu=(1/e)\sqrt{E_FD}$. The electron energy is 50\,eV. We show the total probability (black curves) and the partial probabilities for exciting one-plasmon ($\omega_p$, blue curves) and two-plasmon ($2\omega_p$, red curves) states.}
\label{FigS5}
\end{center}
\end{figure}

Furthermore, we show in Fig.\ \ref{FigS5} the probability for exciting one-plasmon and two-plasmon states under the same trajectories as considered in Figs.\ \ref{FigS3} and \ref{FigS4} for an electron energy of 50\,eV (i.e., $\nu=1.9$).

\section{Acknowledgement}

The author acknowledges helpful and enjoyable discussions with Archie Howie and Darrick E. Chang. He also thanks IQFR-CSIC for providing computers used for numerical simulations. This work has been supported in part by the Spanish MEC (MAT2010-14885) and the EC (Graphene Flagship CNECT-ICT-604391).


\providecommand*{\mcitethebibliography}{\thebibliography}
\csname @ifundefined\endcsname{endmcitethebibliography}
{\let\endmcitethebibliography\endthebibliography}{}

\end{document}